\documentstyle[aps,amssymb,prl,multicol,epsfig]{revtex}
\begin{document}

\author{I.V. Barashenkov \cite{add:igor},
N.V. Alexeeva  \cite{add:nora},
E.V. Zemlyanaya
\cite{add:elena}}
\address{Department of Maths and Applied Maths,
University of Cape Town,  Rondebosch 7701,
South Africa}

\title{Two and three-dimensional oscillons in
nonlinear Faraday resonance}

\maketitle
\begin{abstract}
We study 2D and
3D localised oscillating patterns in a simple
model system exhibiting nonlinear Faraday resonance.
The corresponding amplitude equation is
shown to have exact
soliton solutions which are found to be always unstable
in 3D. On the contrary,
 the 2D solitons are shown to be stable in
 a certain parameter range; hence the damping and
parametric driving are capable of suppressing the nonlinear blowup
and dispersive decay of solitons
in two dimensions.
The negative feedback loop occurs via the enslaving of the soliton's
phase, coupled to the driver,
 to its amplitude and width.
\end{abstract}


\begin{multicols}{2}

Oscillons are localised two-dimensional oscillating
structures which have  recently been detected in
experiments on vertically vibrated
layers of granular material \cite{Swinney},
Newtonian fluids  and suspensions \cite{Faraday,Astruc}.
Numerical simulations  established the existence of stable
oscillons in a variety of pattern-forming systems,
including the Swift-Hohenberg and Ginsburg-Landau
equations, period-doubling maps with continuous spatial coupling,
 semicontinuum theories and hydrodynamic models
\cite{numerical,Astruc}. Although these simulations provided a great
deal of insight into the phenomenology of the oscillons (in particular,
demarcated their existence area on the corresponding
phase diagrams), little is
known about the mechanism by which they acquire or loose their
stability.

 In this Letter, we consider a model equation which
has {\it exact\/} oscillon solutions
and allows an accurate characterisation of their
existence and stability domains.
The main purpose of this work
is to understand how the oscillons manage to
resist the general tendencies toward nonlinearity-induced blow-up or
dispersive decay which are characteristic for localised excitations
in two-dimensional  media.
Our model admits a straightforward generalisation to
three dimensions and we use this opportunity to explore the
existence of stable oscillons in 3D as well.

The model consists of
 a $D$-dimensional lattice of parametrically
  driven nonlinear oscillators (e.g. pendula) \cite{pendula} with
  the nearest-neighbour coupling:
\begin{eqnarray}
\frac{d^2}{d\tau^2} \phi_{\bf k} + \alpha \,
\frac{d}{d \tau} \phi_{\bf k}
  +  2 \kappa  D  \,\phi_{\bf k} -
 \kappa \sum_{|{\bf m} -{\bf k}|=1}
\phi_{\bf m}
 \nonumber \\
+ (1+  \rho  \cos 2 \omega \tau) \sin \phi_{\bf k}=0; \quad
{\bf k}=(k_1,...,k_D).
\label{pendula}
\end{eqnarray}
Assuming that the coupling is strong: $\kappa \gg 1$;
that the damping and driving are weak: $\alpha= \gamma \varepsilon^2$,
$\rho=2h\varepsilon^2$ where $\varepsilon \ll 1$; and that the driving half-frequency is just below
the edge of the linear spectrum  gap: $\omega^2=1-\varepsilon^2$,
the oscillators execute small-amplitude librations of the form
$\phi_{\bf k}= 2 \varepsilon \psi(t, {\bf x}_{\bf k}) e^{-i \omega \tau}
+ c.c. + O(\varepsilon^3)$, where $t=\varepsilon^2 \tau/2$, ${\bf
x}_{\bf k}= \frac{\varepsilon}{\sqrt{\kappa}} {\bf k}$ and the slowly varying amplitude
satisfies
\begin{equation}
i \psi_t + \nabla^2 \psi+ 2|\psi|^2 \psi - \psi = h \psi^* -i \gamma
\psi,
\label{2Dnls}
\end{equation}
the parametrically driven damped nonlinear Schr\"odinger
(NLS) equation.
In 2D,  this equation
was  invoked as a phenomenological model
of nonlinear Faraday resonance in water \cite{Astruc}.
It also
 describes an optical resonator with different
losses for the two polarisation components of the field
\cite{Sanchez}.
In the absence of the damping and driving, all localised
initial conditions in the 2D and 3D NLS
equation are known to either disperse or blow-up in finite time
\cite{Rypdal,Malkin,collapse}.
Surprisingly, numerical simulations of (\ref{2Dnls}) with sufficiently
large $h$ and $\gamma$
revealed the occurrence of stable (or possibly long-lived) stationary
localised excitations \cite{Astruc}.
However  no analytic solutions were found,  and a possible stabilisation
mechanism remained unclear.

In fact there are two exact (though not explicit) stationary
radially-symmetric solutions given by
\begin{equation}
\psi^{\pm}= {\cal A}_{\pm} e^{- i \theta_{\pm}} \, {\cal R}_0({\cal A}_{\pm} r);
\quad (r^2=x_1^2+...+x_D^2),
\label{soliton}
\end{equation}
where
 ${\cal A}_{\pm}^2=1 \pm \sqrt{h^2-\gamma^2}$,
$\theta_+= \frac12 \arcsin(\gamma/h)$, $\theta_-=\frac{\pi}{2}-\theta_+$, and
${\cal R}_0(r)$ is the bell-shaped nodeless solution of
\begin{equation}
\nabla_r^2 {\cal R} - {\cal R} + 2 {\cal R}^3 =0;
\quad
{\cal R}_r(0)={\cal R}(\infty)=0.  \label{master}
\end{equation}
(Below we simply write ${\cal R}$ for ${\cal R}_0$.)
In (\ref{master}), $ \nabla_r^2=\partial_r^2 + (D-1)r^{-1} \partial_r$.
Solutions of Eq.(\ref{master}) in $D=2$ and 3 are well documented in
literature.
(See e.g. \cite{Rypdal} and refs therein.)
One advantage of having an explicit dependence on $h$ and $\gamma$,
is that the  existence domain is characterised by an
explicit formula.
The soliton $\psi^+$ exists for all $h> \gamma$; the $\psi^-$
exists for $\gamma < h < \sqrt{1+ \gamma^2}$.
It is pertinent to add here that for $h< \gamma$,
{\it all} initial conditions decay to zero. This follows from
the rate equation
\begin{equation}
\partial_t |\psi|^2=
\textstyle{\frac{2}{r}} [r(\chi_r |\psi|^2)_r]_r +2 |\psi|^2
(h \sin 2 \chi - \gamma),
\label{rate}
\end{equation}
where   $\psi=|\psi|e^{-i \chi}$.
Defining $N= \int |\psi|^2 d {\bf x}$,
Eq.(\ref{rate}) implies ${N}_t \le 2(h-\gamma)N$ whence
$N(t) \to 0$ as $t \to \infty$.

We now examine the stability of the two solitons.
Linearising Eq.(\ref{2Dnls}) in the small perturbation
\begin{equation}
\delta \psi({\bf x},t)=
e^{(\mu-\Gamma) {\tilde t}-i \theta_{\pm}}[u({\tilde {\bf x}})+i v({\tilde {\bf
x}})],
\label{pert}
\end{equation}
where
${\tilde {\bf x}}={\cal A}_{\pm} {\bf x}$,
${\tilde t}={\cal A}_{\pm}^2 t$,
we get an eigenvalue problem
\begin{equation}
 L_1 u =- (\mu+ \Gamma )v, \quad
(L_0  - \epsilon)v = (\mu- \Gamma )u,
\label{EV_gamma}
\end{equation}
where $\Gamma = \gamma /{\cal A}^2_{\pm}$
and the operators
\begin{equation}
L_0  \equiv - {\tilde \nabla}^2+1 -2 {\cal R}^2({\tilde r}),
\quad
L_1 \equiv L_0 -4 {\cal R}^2({\tilde r}),
\end{equation}
with ${\tilde \nabla}^2= \sum_{i=1}^D \partial^2/\partial {\tilde x_i}^2$.
(We are dropping tildas below.)
The quantity $\epsilon$,
$\epsilon= \pm 2 \sqrt{h^2-\gamma^2}/{\cal A}_{\pm}^2$,
is positive for the $\psi^+$ soliton
and negative for $\psi^-$. Each $\epsilon$ defines a ``parabola"
on the $(h, \gamma)$-plane:
\begin{equation}
h= \sqrt{ \epsilon^2/(2- \epsilon)^2 + \gamma^2}.
\label{h_of_gamma}
\end{equation}
Introducing $\lambda^2=\mu^2-\Gamma^2$
and changing $v({\bf x}) \to (\mu +\Gamma) \lambda^{-1} v({\bf x})$
 \cite{BBK},
Eq.(\ref{EV_gamma}) is reduced to a {\it one-}parameter eigenvalue
problem:
\begin{equation}
(L_0-\epsilon) v= \lambda u, \quad L_1 u= -\lambda v.
\label{EV}
\end{equation}

Since ${\cal R}_0(r)$ is nodeless in $0 \le r < \infty$,
and $L_0 {\cal R}_0=0$, the
operator $L_0-\epsilon$
is positive definite for $\epsilon<0$.
In this case the eigenvalue can be found as a minimum of the
Rayleigh quotient:
\begin{equation}
-\lambda^2=
\min_w \frac{ \langle w| L_1 |w \rangle}
{ \langle w|(L_0-\epsilon)^{-1}|w \rangle}.
\label{min}
\end{equation}
The operator $L_1$ has $D$ zero eigenvalues associated
with the translation eigenfunctions $\partial_i {\cal R}(r)$, $i=1,2,...D$;
hence
it also has a negative eigenvalue with a radial-symmetric eigenfunction
$w_0(r)$. Substituting $w_0$ into the quotient in (\ref{min}), we get
$-\lambda^2<0$ whence $\mu > \Gamma$. Thus the soliton $\psi^-$ is
unstable
(against
a nonoscillatory mode)
for all $D$, $h$
and $\gamma$, and may be safely disregarded.

Before proceeding to the stability of $\psi^+$ (for which
we have $\epsilon>0$),
we make a remark on the
undamped,  undriven case ($\epsilon=0$.)
In 3D, the eigenvalue problem (\ref{EV}) has
 a zero eigenvalue associated with the phase invariance of the
 unperturbed NLS equation (\ref{2Dnls}) and another one, associated with
 the scaling symmetry:
 \begin{equation}
\left( \begin{array}{cc} L_0 & 0 \\ 0 & L_1
\end{array} \right)
\left(\begin{array}{c} {\cal R} \\ -\frac12 (r{\cal R})_r \end{array} \right)=
\left(\begin{array}{c} 0 \\ {\cal R}\end{array} \right).
\label{z1}
\end{equation}
Both the eigenvector $({\cal R},0)^T$ and the rank-2 generalised eigenvector
$(0, -\frac12(r{\cal R})_r)^T$ are radially-symmetric.
In 2D the number of repeated zero eigenvalues associated
with radially-symmetric invariances is four; in addition to those
in (\ref{z1}) we have a two-parameter group of the lens transformations
\cite{Rypdal,Malkin}
giving rise to
 \begin{equation}
\left( \begin{array}{cc} L_0 & 0 \\ 0 & L_1
\end{array} \right)
\left(\begin{array}{c} \frac18 r^2{\cal R} \\ g \end{array} \right)=
\left(\begin{array}{c} - \frac12 (r{\cal R})_r \\
 \hspace*{2mm} \frac18 r^2 {\cal R}\end{array} \right),
\label{z2}
\end{equation}
with some $g(r)$.
When $h^2-\gamma^2$ (or, equivalently, $\epsilon$) deviates from zero,
all the above invariances break down and the two (respectively,
four) eigenvalues move away from the origin on the plane of complex
$\lambda$. The directions of their motion are crucial for the
stability properties.

We can calculate $\lambda(\epsilon)$ perturbatively,
assuming
\begin{eqnarray}
\lambda= \lambda_1 \epsilon^{\frac14}+ \lambda_2 \epsilon^{\frac24}+
\lambda_3 \epsilon^{\frac34} +...,
\label{EV_expansion}
\\
 u= u_1 \epsilon^{\frac14} + u_2 \epsilon^{\frac24}+...,
\quad
v= {\cal R}+ v_1 \epsilon^{\frac14}+ v_2 \epsilon^{\frac24}+...,
\nonumber
\end{eqnarray}
where $v_i=v_i(r)$, $u_i=u_i(r)$. Substituting into
(\ref{EV}), the order  $\epsilon^{1/4}$ gives
 $u_1=-\lambda_1 L_1^{-1} {\cal R}$.
Using (\ref{z1}), $u_1$ is found explicitly:
$u_1=(\lambda_1/2)  (r{\cal R})_r$.
At the order $\epsilon^{2/4}$ we get $u_2=-\lambda_2 L_1^{-1} {\cal R}$
and  equation
$L_0v_2= \lambda_1 u_1$. Since $L_0$ has a null eigenvector, ${\cal R}(r)$,
this equation is only solvable if
 \begin{equation}
 \lambda_1 \int {\cal R}(r) u_1(r) d{\bf x}=
- \lambda_1^2 \, {\textstyle \frac{D-2}{4}}  \int {\cal R}^2(r) d{\bf x} =0.
\label{solva_1}
\end{equation}
In the two-dimensional case
 the condition (\ref{solva_1}) is satisfied for any $\lambda_1$
whereas in $D=3$ we have to set $\lambda_1=0$.
 Next, at the orders
$\epsilon^{3/4}$ and $\epsilon^{4/4}$ we obtain, respectively,
\begin{eqnarray}
L_0 v_3=\lambda_2 u_1 + \lambda_1 u_2= \lambda_1 \lambda_2 (r{\cal R})_r,
\label{34} \\
L_0 v_4= R+ \lambda_1 u_3 + \lambda_2 u_2+ \lambda_3 u_1.
\label{44}
\end{eqnarray}
Eq.(\ref{34}) is solvable both in 2D and 3D.
The solvability condition for (\ref{44}) reduces to
\begin{eqnarray}
\lambda_1^4=
- \frac{ \langle {\cal R}|{\cal R} \rangle}
{ \langle {\cal R}|L_1^{-1} L_0^{-1} L_1^{-1} |{\cal R} \rangle}
=-  16 \frac{ \int {\cal R}^2 d{\bf x}}{\int {\cal R}^2 r^2d{\bf x}},
\label{lambda_1} \\
\lambda_2^2=
 \frac{ \langle {\cal R}|{\cal R} \rangle}
 { \langle {\cal R}| L_1^{-1} |{\cal R} \rangle}
= 4,
\label{lambda_2}
\end{eqnarray}
in two and three dimensions, respectively.

Thus we arrive at two different bifurcation scenarios. In 3D, where
$\lambda_1=0$ and $\lambda_2$ is real,
two  imaginary eigenvalues $\pm|\lambda_2| \epsilon^{1/2}$
 converge at the origin
as $\epsilon \to  0$  from the left.
(This does not mean that the $\psi^-$ soliton
is stable  as  there  still is a pair of finite real
eigenvalues  for $\epsilon<0$.) As $\epsilon$ grows to
positive values, the imaginary pair
$\pm|\lambda_2| \epsilon^{1/2}$ moves onto the real axis. A numerical
study \cite{numerics} of the eigenvalue problem (\ref{EV})
shows that when $\epsilon$ is further increased, the four
 real eigenvalues collide, pairwise, and acquire imaginary parts.
 Importantly,  for all $0< \epsilon <1$ the imaginary
 parts remain smaller in magnitude than the real parts. This means that
 ${\rm Re\/} \mu$ remains
 greater than $\Gamma$ all the time,
implying that
the three-dimensional
  $\psi^+$ soliton is unstable for all $h$ and $\gamma$.

The bifurcation occurring in 2D is more unusual. As $\epsilon$
approaches zero
from the left, {\it four\/} eigenvalues converge at the origin, two
along the real and two along  imaginary axis: $\lambda
\approx \pm |\lambda_1| (-\epsilon)^{1/4},
\pm i|\lambda_1| (-\epsilon)^{1/4}$. As $\epsilon$ moves to positive, the four eigenvalues
start diverging at $45^\circ$ to the real and imaginary axes.
Hence to the leading order,
${\rm Im\/} \lambda \approx {\rm Re\/} \lambda$,
and in order to make
a conclusion about the stability, we need to calculate the
higher-order corrections.  The order $\epsilon^{5/4}$ produces
a solvability condition
\[
\lambda_1^3 \lambda_2  \langle {\cal R}|L_1^{-1} L_0^{-1} L_1^{-1}
|{\cal R} \rangle=
\frac{\lambda_1^3 \lambda_2}{16} \int {\cal R}^2r^2 d{\bf x}= 0,
\]
 yielding $\lambda_2=0$. (Here we made use of  (\ref{z2}).)
 Finally, the order $\epsilon^{6/4}$
defines $\lambda_3$ (where $g(r)$ is as in (\ref{z2})):
\begin{figure}
\begin{center}
\psfig{file=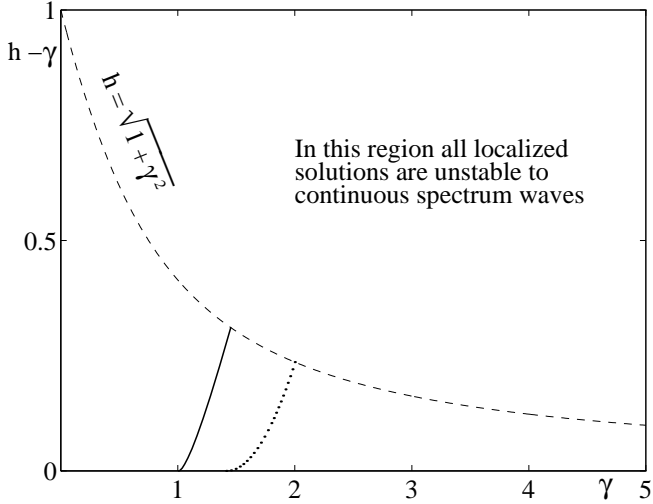,width=1.\linewidth}
\end{center}
\caption{\sf Stability diagram for two-dimensional solitons.
The $(\gamma, h-\gamma)$-plane is
used for visual clarity. No localised or periodic attractors
exist for $h< \gamma$ (below the horisontal axis).  The region of
stability of the soliton $\psi^+$  lies  to the right of the
solid curve.
The dotted curve
 gives the variational approximation
to the stability boundary of the  $\psi^+$
soliton: $h=(1+ \gamma^4)^{1/2}$,
$\gamma \ge \sqrt{2}$.}
\label{chart}
\end{figure}
\begin{equation}
\lambda_3= \frac{1}{\lambda_1} +
\frac{\lambda_1^3}{2} \frac{\int g(r) {\cal R}(r) r^2 d{\bf x}}
{\int {\cal R}^2(r) r^2 d{\bf x}}.
\label{lambda_3}
\end{equation}

Taking $\lambda_1$ in the first quadrant, $\lambda_1= e^{i \pi/4}
|\lambda_1|$, and doing the integrals in (\ref{lambda_1}),
(\ref{lambda_3})
numerically, we conclude that $\lambda_3$ is in the second quadrant,
$\lambda_3=e^{3i\pi/4} |\lambda_3|$, which implies that $|{\rm Im\/}
\lambda|> |{\rm Re\/} \lambda|$. In terms of
$\lambda$, the stability criterion ${\rm Re\/} \mu \le \Gamma$ is
written as $\gamma \ge \gamma_c$, where
 \begin{equation}
\gamma_c(\epsilon) \equiv \frac{2}{2-\epsilon} \cdot \frac{{\rm Re\/} \lambda(\epsilon)
  \, {\rm Im\/} \lambda(\epsilon)}
 {\sqrt{({\rm Im\/} \lambda)^2- ({\rm Re\/} \lambda)^2}}.
 \label{gce}
 \end{equation}
The smallest $\gamma$ for which the soliton can be stable, is given by
   \begin{equation}
\lim_{\epsilon  \to 0} \gamma_c(\epsilon)=
 \frac{1}{2\sqrt{2}}  |\lambda_1|^{3/2} |\lambda_3|^{-1/2}.
 \label{gc0}
 \end{equation}
Substituting for $\lambda_1$, $\lambda_3$ their numerical values,
(\ref{gc0}) gives $\gamma_c(0)=1.00647$. For $\epsilon \neq 0$
we obtained $\lambda(\epsilon)$ by solving the eigenvalue problem
(\ref{EV}) directly \cite{numerics}. Here we
have restricted ourselves to
radially-symmetric $u(r)$ and $v(r)$. Expressing $\epsilon$ via
$\gamma_c$ from (\ref{gce}) and feeding into (\ref{h_of_gamma}),
we get the stability
boundary on the $(h, \gamma)$-plane (Fig.1).

Asymmetric perturbations do not lead to any instabilities in 2D.
To show this, we factorise, in (\ref{EV}),
$u({\bf x})={\tilde u}(r) e^{i m \varphi}$ and
$v({\bf x})={\tilde v}(r) e^{i m \varphi}$,
where $\tan \varphi= y/x$
and $m$ is an integer. The eigenproblem (\ref{EV})
remains the same, with only the operators $L_0$ and $L_1$
being replaced by
\begin{equation}
L_0^{(m)}
 \equiv L_0
  +{m^2}/{r^2},
\quad
L_1^{(m)} \equiv L_1+ {m^2}/{r^2}.
\label{m2r2}
\end{equation}
 The crucial observation
now is that $L_0^{(m)}$ with $m^2 \ge 1$ does not have {\it any\/}
(not even positive)
discrete eigenvalues.
We verified this numerically for $m^2=1$; this rules out their
appearance for all other $m$. Therefore the operator
$L_0^{(m)}-\epsilon$ with $\epsilon <1$ is positive definite,
and the eigenvalues of the problem (\ref{EV}) can be found from
the variational principle (\ref{min}). The operator $L_1^{(1)}$
has a zero eigenvalue with the eigenfunction $w^{(1)}(r)={\cal
R}_r(r)$ which has no nodes for $0<r<\infty$; hence its
all other eigenvalues (if exist) are positive. This also implies
that $L_1^{(m)}$ with $m^2 >1$ are positive definite. Thus the
minimum of the Rayleigh quotient (\ref{min}) is zero for
$m^2=1$ and positive for $m^2 >1$.

Besides the nodeless solution ${\cal R}_0(r)$, the ``master"
equation (\ref{master}) has solutions ${\cal R}_n(r)$ with $n$
nodes, $n=1,2,...$. These give rise to a sequence of nodal
solutions of the damped-driven NLS (\ref{2Dnls}), defined
by Eq.(\ref{soliton}) with ${\cal R}_0 \to {\cal R}_n$.
It is easy to realise that the solitons $\psi_n^-$ are unstable
against radially-symmetric nonoscillatory modes for
 all $h$, $\gamma$, $n$ and $D$. (The proof is a simple
generalisation of the one for $\psi_0^-$.)
To examine the stability of the
$\psi^+_{n}$ soliton, we  solved the eigenvalue problem (\ref{EV})
numerically,
with the operators $L_{0,1}^{(m)}$ as in
(\ref{m2r2}).
In 3D, positive real eigenvalues
(with radially-symmetric eigenfunctions) are present in the
spectrum for all
$\epsilon$;
thus the three-dimensional nodal solitons are always prone to a
symmetric collapse or dispersive spreading.
In 2D, the $\psi_n^+$ solitons are
stable against radially-symmetric
perturbations  for sufficiently large $\gamma$.
However, these solutions turn out to be always unstable
against azimuthal perturbations. In particular, the $\psi_1^+$ soliton
has instabilities associated with $1 \le m \le 5$, and the
$m=4$ mode has the largest growth rate for all $\epsilon$.
The corresponding
eigenvalue $\lambda$ is real and the eigenfunctions
$u(r)$ and $v(r)$ have a single maximum near the position
of the lateral minimum of the function $R_1(r)$.
Following Ref.\cite{Akhmediev} where a similar scenario was
described for  nodal waveguides in a
saturable
self-focusing medium, the above observation suggests that
the $\psi_1^+$
soliton will decay into 5  solitons $\psi_0^+$:
one at the origin and four others placed symmetrically around it.
Next, the $\psi_2^+$ solution has azimuthal instabilities
with $1 \le m \le 10$. The analysis of the corresponding
eigenfunctions suggests that, depending on
$h$ and $\gamma$, the decay products will comprise 11 to 13
 $\psi_0^+$-solitons: 1 at the origin; 3 or 4
 placed symmetrically around
it; and 7 or 8 forming an outer ring.
We verified these predictions via direct numerical simulations
of the time-dependent array (\ref{pendula}); the simulations
corroborated the above scenario.
Thus the nodal solutions can be interpreted
as degenerate coaxial complexes of the nodeless solitons and
serve as nuclei of
symmetric multisoliton patterns.

Lastly, we need to  understand the stabilisation mechanism
in qualitative terms.
To this end, we use the variational approach.
The equation (\ref{2Dnls}) is
derivable from the stationary action principle with the
Lagrangian
\[
{\cal L}= e^{2 \gamma t} {\rm Re\/} \int  (
i \psi_t \psi^* -|\nabla \psi|^2 -|\psi|^2
+ |\psi|^4 -h \psi^2 ) d {\bf x}.
\]
Choosing the ansatz  $\psi=
\sqrt{A}  e^{-i \theta -(B+i \sigma) r^2}$ \cite{ansatz,1dlonghi}
 with $A,B,\theta,\sigma$ functions of $t$,
 this reduces, in 2D, to
\begin{eqnarray}
{\cal L}=e^{2 \gamma t}  \frac{A}{B}
\left[
{\dot \theta}-1  +  \frac{\dot \sigma}{2B}
- \frac{2B}{\cos^2 \phi} + \frac{A}{2} \right. \nonumber \\
 \left. \phantom{\frac{A}{B}} - h \cos (\phi +2 \theta)
\cos \phi \right]; \quad \tan \phi= \sigma/B.
\label{Lagrangian}
\end{eqnarray}
The 4-dimensional dynamical system defined
by (\ref{Lagrangian}), has two stationary points
representing  the $\psi^{\pm}$ solitons. In
agreement with the stability properties of the
solitons in the full PDE,  the $\psi^+$ stationary
point is unstable
 for small $\gamma$ but stabilises for larger dampings  (Fig.\ref{chart}).
When $\gamma$ is large we can expand
$A=A_0+\frac{1}{\gamma} A_1+...,
B= B_0+ \frac{1}{\gamma} B_1 $ $+...$,
$\theta=\frac{\pi}{4} +\frac{1}{\gamma} \theta_1+...$,
$\sigma=\frac{1}{\gamma} \sigma_1+...$.
Letting $h=\gamma+
\frac{c}{2\gamma}$ where $0 \le c \le 1$,
defining $T=\frac{t}{\gamma}$
and matching coefficients of like powers of
$\frac{1}{\gamma}$, yields a  2-dimensional  system
\begin{eqnarray}
d A_0/dT = A_0[c+ 8 \sigma_1 -4\theta_1^2 + 2(\sigma_1/B_0)^2],
 \label{A} \\
 d B_0/dT = 8 \sigma_1B_0 +4 \sigma_1\theta_1+4(\sigma_1^2/B_0),
\label{B} \\
\theta_1= \textstyle{\frac12}+2B_0-\textstyle{\frac34} A_0, \quad
 \sigma_1=\textstyle{\frac12} A_0B_0-2B_0^2
 \label{th_si}.
 \end{eqnarray}
Like their parent system (\ref{Lagrangian}),
Eqs.(\ref{A})-(\ref{th_si}) have two fixed points, the saddle at
$B_0^-=\frac12-\sqrt{c}$, $A_0^-=4B_0^-$ and a stable focus at
$B_0^+=\frac12 +\sqrt{c}$, $A_0^+=4B_0^+$.

According to (\ref{rate}), the  soliton's phase
$\chi=\theta+ \sigma r^2$ controls the creation and
annihilation  of the soliton's elementary constituents
(whose density is $|\psi|^2$).
(If Eq.(\ref{2Dnls}) is used as a model equation for Faraday
resonance in granular media or fluids, $\int |\psi|^2$ d{\bf x} has the
meaning of the number of grains or mass of the fluid captured in
the oscillon.) Since the creation and annihilation
occurs mainly in the core of the soliton, the variable phase component
$\sigma r^2$ plays a marginal role in this process.
 Instead, the significance of the quantity
 $\sigma$  is in that it controls the flux of the constituents
  between the core and the periphery of
 the soliton --- see the $\chi_r$-term in the r.h.s. of (\ref{rate}).

 If we perturb
the stationary point $\psi^+$  in the 4-dimensional phase space of
(\ref{Lagrangian}),
the variables $\theta$ and $\sigma$ will zap,
  within
a very short time $\Delta t \sim \frac{1}{\gamma}$,
onto the 2-dimensional
subspace defined by the constraints (\ref{th_si}). After this short
transient
the evolution of $\theta$ and $\sigma$ will be immediately following
 that of
the soliton's amplitude
   $\sqrt{A}$ and width $1/\sqrt{B}$.
 In the case of the $\psi^+$ soliton,
 this provides a negative feedback: perturbations in $A$ and $B$
  produce only such changes in the phase and flux  that  the new
  values of $\theta$ and $\sigma$
 stimulate the recovery of the stationary values of $A$ and $B$.
 (The phase $\theta$ works to restore the number of constituents
 while $\sigma$ rearranges them within the soliton.)
 In the case of the $\psi^-$ the feedback is positive:
 the perturbation-induced phase and flux (\ref{th_si}) strive to amplify the
 perturbation of the soliton's amplitude and width
 still further. Finally, for small
 $\gamma$ the
 coupling of
 $\theta$ and $\sigma$ to $A_0$ and $B_0$ is via differential rather
 than
 algebraic equations. In this case
 the dynamics of the phase and flux is inertial
 and their changes may not catch up with those of the
 amplitude and
 width. The feedback loop is destroyed and the soliton destabilises.

We thank Dominique Astruc, Sergei Flach,
Stephano Longhi,
and Dmitry Pelinovsky for useful discussions.
The work of
E.Z.  supported
by  RFBR
grant 0001-00617.

\end{multicols}


\begin{thebibliography}{99}
\bibitem{numerics}
For the solution of  Eq.(\ref{master})
we used a  fourth-order Newtonian iteration, on the intervals
(0,30) and (0,45), with the stepsize $\Delta r=2.5 \cdot
10^{-3}$ and $1.25 \cdot 10^{-3}$.
The eigenvalue problem (\ref{EV}) was solved by the Fourier expansion
over 200 modes (and verified with 400 and 800 modes.)
\end{thebibliography}
\end{document}